\documentclass[reprint,amsmath,amssymb,aps]{revtex4-1}

\usepackage{graphicx}
\usepackage{dcolumn}
\usepackage{bm}
\usepackage{hyperref}
\usepackage[mathlines]{lineno}

\usepackage{amsmath,amssymb}  
\usepackage{dsfont}

\newcommand{\bra}[1]{\langle#1|}
\newcommand{\ket}[1]{|#1\rangle}

\DeclareMathOperator{\Tr}{Tr}

\begin{document}

\preprint{APS/123-QED}

\title{Appearance of causality in process matrices when performing fixed-basis measurements for two parties}



\author{Veronika Baumann}
\affiliation{Faculty of Informatics, Universit\`a della Svizzera italiana, Via G. Buffi 13, CH-6900 Lugano, Switzerland}
\affiliation{Faculty of Physics, University of Vienna, Boltzmanngasse 5, 1090 Vienna, Austria}
\affiliation{Institute for Quantum Optics and Quantum Information, Boltzmanngasse 3, 1090 Vienna, Austria}
\author{\v Caslav Brukner}
\affiliation{Faculty of Physics, University of Vienna, Boltzmanngasse 5, 1090 Vienna, Austria}
\affiliation{Institute for Quantum Optics and Quantum Information, Boltzmanngasse 3, 1090 Vienna, Austria}

\date{\today}

\begin{abstract}
The recently developed framework for quantum theory with no global causal order allows for quantum processes in which operations in local laboratories are neither causally ordered nor in a probabilistic mixture of definite causal orders. The causal relation between the laboratories is described by the process matrix. We show that, if the inputs of the laboratories are measured in a fixed basis, one can introduce an effective process matrix which is operationally indistinguishable from the original one. This effective process matrix can be obtained by applying the von Neumann-L{\" u}ders update rule for nonselective measurements to the original process matrix and in the bipartite case it is compatible with a definite causal order. The latter extends the original Oreshkov et al. proof where one considers that both the measurement of the input and the re-preparation of the output are performed in a fixed basis. 
\end{abstract}

\maketitle


{\em Introduction:} In all our theories, including quantum field theory on curved space-time, causal structure is fixed. However, this is unlikely to be present in a theory unifying quantum physics and general relativity, where the space-time geometry is expected to be both dynamical, as in general relativity, as well as indefinite, due to quantum theory \cite{hardy2005probability}. Recently, frameworks for quantum theory have been developed that, unlike the standard formulations, do not presume a fixed background causal structure \cite{hardy2005probability, hardy2011reformulating, chiribella2009theoretical, Oreshkov2012,leifer2013towards}. Moreover, the ``process framework'' \cite{Oreshkov2012} was shown to allow for processes in which two operations are neither causally ordered nor in a probabilistic mixture of definite causal orders. 

The central object of the process framework is the ``process matrix'', which describes causal relations  between local laboratories. Each such laboratory features a device with an input and an output. At every run of the experiment, the laboratory is opened for the input to enter. Then a quantum operation (a complete-positive map) is performed while the laboratory is closed, after which the output of the operation is sent away. For example, the operation can consist of a measurement of the input and then re-preparation of the output. 

A bipartite process matrix $W$ is called causally separable if it can be decomposed as a convex combination of causally ordered ones: $W= pW^{B\preceq A} + (1-p)W^{A\preceq B}$, where $0\leq p \leq 1$ and $W^{A \preceq B}$ is a process with which Bob cannot signal to Alice and $W^{B \preceq A}$ a process with which Alice cannot signal to Bob. An explicit example of a causally non-separable process is the ``quantum switch" where the causal structure is in quantum superposition \cite{chiribella2012perfect,chiribella2013quantum,araujo2015witnessing,oreshkov2015causal}. Moreover, certain causally non-separable processes can violate ``causal inequalities'' which are satisfied by all correlations that are unidirectional signaling (either from $A$ to $B$, or from $B$ to $A$) or non-signaling or can be decomposed in a classical (convex) mixture of such correlations \cite{branciard2015simplest}.  

In ref. \cite{Oreshkov2012} it was shown that in the bipartite case, if the inputs are measured in a fixed basis {\em and} the outputs are reprepared in a fixed basis, the causal relations between the local operations are compatible with a global causal order. Hence, the {\em effective process matrix} the parties can reconstruct from their data under this restricted set of operations is always causally separable. 
This is not true for three or more parties where there exist classical process matrices (i.e. where bases are fixed for the input measurement and the output preparation) that violate causal inequalities \cite{baumeler2014maximal}. 

Here we show that the causal relations between two local laboratories are compatible with a global causal order if {\em only} the inputs are measured in a fixed basis, thus relaxing the assumption of the original proof \cite{Oreshkov2012}. The effective process matrix $W_{eff}$ can be obtained by applying the von Neumann-L{\"u}ders update rule \cite{luders1950zustandsanderung,neumann1932mathematische} for non-selective measurements to the original $W$ matrix. $W \rightarrow \sum_{m,n} P_{n}\otimes P_{m} W P_{n} \otimes P_{m} \equiv W_{eff}$, where $P_{n}$ and $P_{m}$ are projectors of the inputs of $A$'s and $B$'s laboratory onto the respective measurement basis. Moreover, $W_{eff}$ is causally separable. This is analogous to the situation where an entangled state is transformed into a separable one, upon a non-selective measurement in a product basis. The observed correlations are then compatible with the separable state $\rho_{sep}$ that can be obtained from the original $\rho$ by $\rho \rightarrow \sum_{m,n} P_{n}\otimes P_{m} \rho P_{n}\otimes P_{m} \equiv \rho_{sep}$, where $P_{n}$ and $P_{m}$ are orthogonal projectors on the subspaces of $A$ and $B$.


{\em Process formalism:} We consider causal relations between two laboratories. In the laboratories quantum mechanics with a fixed (local) causal structure holds. The laboratories receive inputs $\in \mathcal{L}(\mathcal{H}^{X_1})$ and produce outputs $\in \mathcal{L}(\mathcal{H}^{X_2})$, $X=A, B$, but are otherwise isolated. Each party applies a complete-positive map from a set \{$\mathcal{M}_k^{X}$\}, which is indexed by the observed outcome $k$. The joint probability for a pair of maps, ($\mathcal{M}_i^{A},\mathcal{M}_j^{B})$, is given by a generalized Born rule \cite{Oreshkov2012}
\begin{equation} \label{PCJmatrix}
P(\mathcal{M}_i^{A},\mathcal{M}_j^{B}) = \Tr \left[ W ( M_i^{A_1 A_2} \otimes M_j^{B_1 B_2}) \right],
\end{equation}
where $M_k^{X_1 X_2} $ are the Choi-Jamiolkowski (CJ) matrices \cite{choi1975completely, jamiolkowski1972linear} of the operations $\mathcal{M}_k^{X}$ and $W  \in \mathcal{L} (\mathcal{H}^{A_1} \otimes \mathcal{H}^{A_2} \otimes \mathcal{H}^{B_1} \otimes \mathcal{H}^{B_2})$ is the process matrix. Summing over all outcomes of any party should give a complete-positive, trace preserving map, which means the CJ matrices satisfy $\Tr_{X_2}(\sum_k M_k^{X_1 X_2}) =\mathds{1}_{X_1}$. Requiring that probabilities are positive and sum up to one (where the parties are allowed to share entangled states) implies $W\geq0$ and a certain form of $W$, which automatically excludes causal loops,   
\begin{equation}\label{Wstruct}
W = \frac{1}{d} \left( \mathds{1}^{A_1A_2 B_1 B_2} + \sum_{i} c_i W_i^{A_1A_2 B_1 B_2} \right).
\end{equation}
Here $d=dim(\mathcal{H}^{A_1})\cdot dim(\mathcal{H}^{B_1})$, $c_i \in \mathbb{R}$ and $W_i^{A_1A_2 B_1 B_2}$ are traceless and have trivial entry in at least one of the output spaces and nontrivial entry in the respective input space, i.e they are either of $w_i^{A_1A_2B_1}\otimes \mathds{1}^{B_2}$ or $w_i^{B_1B_2A_1}\otimes \mathds{1}^{A_2}$ form. For example, a channel from Alice to Bob is described by a process matrix of the form $W=w^{A_1A_2B_1} \otimes \mathds{1}^{B_2}$, while one from Bob to Alice by $W=w^{B_1B_2A_1} \otimes \mathds{1}^{A_2}$. Equations \eqref{PCJmatrix} and \eqref{Wstruct} can readily be generalized to more parties.


{\em Effective process matrix:} The process matrix encodes causal relations between local laboratories. If local operations are restricted to a set, one can find an {\em effective process matrix} $W_{eff}$ that returns the same joint probabilities as the original process matrix $W$: 
\begin{equation} \label{Weffdef}
\begin{aligned}
& \Tr\left[W (M^{A_1A_2}_i \otimes M^{B_1B_2}_j)\right] \overset{!}{=} \\
&\Tr\left[W_{eff} (M^{A_1A_2}_i \otimes M^{B_1B_2}_j)\right],
\end{aligned}
\end{equation} 
for all $i,j$ from the set. One might think of $W_{eff}$ as representing the causal structure observers in closed laboratories would deduce from the correlations between them, when their operations are limited. The idea of an effective process matrix can straightforwardly be generalized to many local laboratories. 

An example of restricted operations are classical ones for which all CJ matrices are diagonal both in the input and the output space, $M^{A_1A_2}_i= \sum_{n,r} p(i,r|n) \ket{n}\bra{n}^{A_1} \otimes \ket{r}\bra{r}^{A_2}$ and $M^{B_1B_2}_j= \sum_{m,s} p(j,s|m) \ket{m}\bra{m}^{B_1} \otimes \ket{s}\bra{s}^{B_2}$, where $p(k,t|l)$ is the probability of outcome $k$ and re-preparation of state $t$ given input $l$. (In what follows sums in expressions of process matrices are always taken over all appearing indices). Given a general process matrix $W=  \sum w_{nn'rr'mm'ss'}  \ket{n}\bra{n'}^{A_1}\otimes \ket{r}\bra{r'}^{A_2}\otimes  \ket{m}\bra{m'}^{B_1}\otimes \ket{s'}\bra{s'}^{B_2}$, the effective process matrix is then also diagonal $W_{eff} =  \sum w^{eff}_{nrms}  \ket{n}\bra{n}^{A_1}\otimes \ket{r}\bra{r}^{A_2}\otimes  \ket{m}\bra{m}^{B_1}\otimes \ket{s}\bra{s}^{B_2}$, with coefficients $w^{eff}_{nrms}=w_{nnrrmmss}$. 
While in the bipartite case $W_{eff}$ has been proven to be causally separable~\cite{Oreshkov2012}, this is no longer true for more than two parties ~\cite{baumeler2014maximal}.

We will now show that for the bipartite case the effective process matrices that are diagonal {\em only in the input spaces} of the two parties are still causally separable. 
We consider effective process matrices of the form  
\begin{equation} \label{Weff2basis}
\begin{aligned}
&W_{eff}=
\sum w^{eff}_{ni i' m j j'} \\
&
\ket{n}\bra{n}^{A_1}\otimes \ket{i}\bra{i'}^{A_2}\otimes  \ket{m}\bra{m}^{B_1}\otimes \ket{j}\bra{j'}^{B_2}, 
\end{aligned}
\end{equation}
where \{$\ket{i}$\} and \{$\ket{j}$\} are arbitrary orthonormal bases of $\mathcal{H}^{A_2}$ and $\mathcal{H}^{B_2}$ and $w^{eff}_{nii'mjj'}=w_{nnii'mmjj'}$ analogous to the classical case.
 As any two-party process matrix, \eqref{Weff2basis} can be written as
\begin{equation} \label{Wstart}
\begin{aligned}
W_{eff} = \frac{1}{d} ((1+\lambda_0) \mathds{1} + \kappa_1+\kappa_2), 
\end{aligned}
\end{equation}
where $\lambda_0\in [ -1,0]$ is the minimal eigenvalue of $\sum_{i} c_i W_i^{A_1A_2 B_1 B_2}$, $\kappa_1$  acts trivially on $\mathcal{H}^{B_2}$, $\kappa_2$ acts trivially on $\mathcal{H}^{A_2}$ and $\kappa_1+\kappa_2 \geq 0$. 
In the original proof (see supplementary information of ref. \cite{Oreshkov2012}), one makes use of terms $P_{(n,m)}= \ket{n} \bra{n}^{A_1} \otimes \mathds{1}^{A_2} \otimes \ket{m} \bra{m}^{B_1} \otimes \mathds{1}^{B_2}$, where \{$\ket{n} \bra{n}$\} and \{$\ket{m} \bra{m}$\} form orthogonal sets. These terms are added to and subtracted from $\kappa_1$ and $\kappa_2$ for every pair $(n,m)$ in a systematic way,  such that the sum $\kappa_1+\kappa_2$ (and hence $W_{eff}$) remains  unchanged. 
In that way, one can produce positive eigenvalues and arrive at matrices $\overline{\kappa}^{A_1 A_2 B_1} \geq 0$ and $\overline{\kappa}^{A_1 B_1 B_2} \geq 0$ (the superscript denotes spaces with non trivial entries) such that $(1+\lambda_0) \mathds{1}+\kappa_1+\kappa_2 =\overline{\kappa}^{A_1 A_2 B_1}+\overline{\kappa}^{A_1 B_1 B_2}$. Hence $W_{eff} = \frac{1}{d} (\overline{\kappa}^{A_1 A_2 B_1} + \overline{\kappa}^{A_1 B_1 B_2}) = p W^{A\preceq B} + (1-p) W^{B\preceq A}$  where $p=\Tr(\overline{\kappa}^{A_1 A_2 B_1})/d'$ with $d'=d \cdot dim(\mathcal{H}^{A_2})\cdot dim(\mathcal{H}^{B_2})$ and $1-p=\Tr( \overline{\kappa}^{A_1 B_1 B_2})/d'$.

Examining the proof in detail shows that the systematic modification of the eigenvalues requires\\ 
\vspace{-1em}
\begin{equation} \label{condi1}
[\kappa_1, \kappa_2]= [\kappa_1, P_{(n,m)}]=[P_{(n,m)}, \kappa_2]=0
\end{equation} 
and the joint eigenvectors of $\kappa_1, \kappa_2$ and $ P_{(n,m)}$ to be product vectors 
\begin{equation} \label{condi2}
\ket{\psi^{(n,m)}}=\ket{n}^{A_1} \otimes \ket{a^{(n,m)}}^{A_2} \otimes \ket{m}^{B_1} \otimes  \ket{b^{(n,m)}}^{B_2}
\end{equation}
with $\{\ket{n} \}$ and $\{\ket{m} \}$ being orthonormal bases of the input spaces $\mathcal{H}^{A_1}$ and $\mathcal{H}^{B_1}$. For every pair $(n,m)$ the $\{\ket{a^{(n,m)}} \}$, $\{\ket{b^{(n,m)}} \}$ are orthogonal bases of $\mathcal{H}^{A_2}$ and $\mathcal{H}^{B_2}$ respectively. Together \eqref{condi1} and \eqref{condi2} ensure that $m(n,a,m,b) = m_1(n,a,m)+m_2(n,m,b)$, where $m(\cdot)$ is the eigenvalue of $\kappa_1+\kappa_2$ and $m_1(\cdot)$, $m_2(\cdot)$ are the eigenvalues of $\kappa_1$, $\kappa_2$ respectively.
For process matrices \eqref{Weff2basis} the commutation relations are trivially fulfilled. Moreover, as \eqref{condi2} are eigenvectors of $ P_{(n,m)}$ one has $\ket{\psi^{(n,m)}}=\sum_{jl} c_{jl} \ket{n}^{A_1} \otimes \ket{j}^{A_2} \otimes \ket{m}^{B_1} \otimes  \ket{l}^{B_2}$. 
For a given $(n,m)$ any $\kappa_1$ from \eqref{Weff2basis} acts on $\ket{\psi^{(n,m)}}$ as 
$\mathds{1}^{A_1}\otimes A_{(n,m)}^{A_2}\otimes \mathds{1}^{B_1 B_2}$
and analogously 
$\kappa_2$ as $\mathds{1}^{A_1 A_2 B_1}\otimes B_{(n,m)}^{B_2}$, with hermitian matrices $A_{(n,m)}$ and $B_{(n,m)}$.
Therefore, $\ket{\psi^{(n,m)}}=\ket{n}^{A_1} \otimes \ket{a^{(n,m)}}^{A_2} \otimes \ket{m}^{B_1} \otimes  \ket{b^{(n,m)}}^{B_2}$, where $\ket{a^{(n,m)}}$ and $ \ket{b^{(n,m)}}$ are eigenvectors of $A_{(n,m)}^{A_2}$ and $B_{(n,m)}^{B_2}$.
Note that conditions \eqref{condi1} and \eqref{condi2} are sufficient for causal separability but not necessary. To see this, simply consider 
$
W_{0}= \frac{p}{d}(\mathds{1}-\sigma_z^{A_1}\otimes\sigma_z^{A_2}\otimes \sigma_x^{B_1})$ $
+\frac{1-p}{d}(\mathds{1}+\frac{1}{2}\sigma_z^{A_1}\otimes \sigma_x^{B_2} +\frac{1}{2}\sigma_x^{A_1}\otimes\sigma_x^{B_1}\otimes \sigma_z^{B_2})
$.
The two nontrivial terms do not commute. Moreover no nontrivial $P_{(n,m)}$ exists, which commutes with both terms, since it would have to commute with both $\sigma_x$ and $\sigma_z$. Hence neither \eqref{condi1} nor \eqref{condi2} is fulfilled, although $W_{0}$ is causally separable per definition.

{\em Interpretation:} We considered bipartite local operations that consist of a  measurement of the input and a re-preparation of the output. The CJ representation of the complete-positive map has the form  $M^{X_1X_2}_{\phi_1 \phi_2} = \ket{\phi_1}\bra{\phi_1}^{X_1}\otimes \ket{\phi_2}\bra{\phi_2}^{T \; X_2}$, where $\ket{\phi_1}\bra{\phi_1}$ describes the measured input state and $\ket{\phi_2}\bra{\phi_2}$ is the prepared output state. Following Holevo's classification~\cite{holevo1998quantum} of entanglement breaking quantum channels, one can denote maps, where both the measurement and the re-preparation are performed in a fixed basis, as classical-classical maps. A classical-quantum map then corresponds to an operation in which the input is measured in a fixed basis, but the output is arbitrary. The CJ representation of such map is $M^{X_1X_2}_i= \sum_{n} p(i|n) \ket{n}\bra{n}^{X_1} \otimes \rho_i^{X_2}$, where $p(i|n)$ is the probability to prepare state $\rho_i^{X_2}$ given the input is measured to be in state $\ket{n}$. 

 If the local operations are restricted to  classical-quantum maps, the effective process matrix~\eqref{Weff2basis} is causally separable. Moreover, it is related to the original process matrix through the update rule
\begin{equation}
W \rightarrow W_{eff} = \sum_{n,m} P_{(n,m)} W P_{(n,m)},
\end{equation}
where $P_{(n,m)}= P_n \otimes P_m$, which is analogous to the von Neumann-L{\"u}der rule for the state update under a non-selective measurement. 

Consider the projection of a general state upon a measurement onto a product basis states. While the original bipartite state $\rho$ might be entangled and contain quantum correlations, the state after the measurement $\sum_{n,m} P_{n}\otimes P_{m} \rho P_{n}\otimes P_{m}$ is separable and contains only classical correlations.
Analogously, while the original $W$ might encode  correlations with no definite causal order, after the measurements are performed in a fixed basis, the process matrix is causally separable and the
correlations are compatible with a global causal order.

Note that there is no analogy to the update rule for {\em selective} measurements. In general, $P_{(n,m)} W P_{(n,m)}$ is not a valid process matrix. It contains terms that correlate the output spaces of Alice and Bob, or  the input and output space of a single party and is therefore not of the form \eqref{Wstruct}.

In processes violating causal inequalities the causal order cannot be assumed to be fixed prior to and independent of local operations. However, by choosing the measurement basis for the inputs, local observers are able to (non-selectively) ``project'' the initial process matrix to a causally separable one -- a desired (noisy) channel between them. In general, this channel will be a convex mixture of a channel from $A$ to $B$ and a channel from $B$ to $A$. This provides a way to interpret violation of causal inequalities by processes.\\

\begin{acknowledgments}
This work has been supported by the European Commission Project RAQUEL and the Austrian Science Fund (FWF) through SFB FoQuS, and the Individual Project 2462. We also acknowledge the support of the Swiss National Science Foundation (SNF) and the National Centre of Competence in Research Quantum Science and Technology (QSIT).
\end{acknowledgments}


\bibliography{NewOne}
\end{document}